\documentclass[
  aps,
  prl,
  reprint,
  amsmath,amssymb
]{revtex4-2}

\usepackage[english]{babel}
\usepackage{bbold}
\usepackage{bbm}
\usepackage{color}
\usepackage{rotating}
\usepackage{graphicx}
\usepackage{multirow}
\usepackage{color}
\usepackage{array}

\usepackage{xcolor}
\definecolor{RoyalBlue}{HTML}{4169e1}
\definecolor{ForestGreen}{HTML}{228b22}

\usepackage[colorlinks = true,
            linkcolor = blue,
            urlcolor  = blue,
            citecolor = blue,
            anchorcolor = blue]{hyperref}

\usepackage{comment}

\begin{document}
\title{Tuning Universality in Deep Neural Networks}

\author{Arsham Ghavasieh}
\email{arsham.ghavasieh@gmail.com}
\affiliation{Center for Complex Networks and Systems Research, Luddy School of Informatics, Computing, and Engineering, Indiana University, Bloomington, Indiana 47408, USA}

\date{\today}

\begin{abstract}
Deep neural networks (DNNs) exhibit crackling-like avalanches whose origin lacks a mechanistic explanation. Here, I derive a stochastic theory of deep information propagation (DIP) by incorporating Central Limit Theorem (CLT)-level fluctuations. Four effective couplings $(r, h, D_1, D_2)$ characterize the dynamics, yielding a Landau description of the static exponents and a Directed Percolation (DP) structure of activity cascades. Tuning the couplings selects between avalanche dynamics generated by a Brownian Motion (BM) in a logarithmic trap and an absorbed free BM, each corresponding to a distinct universality classes. Numerical simulations confirm the theory and demonstrate that activation function design controls the collective dynamics in random DNNs.
\end{abstract}

\maketitle

Statistical physics has long provided a framework for understanding collective neural dynamics~\cite{hopfield1982neural, cossart2003attractor, schneidman2006weak, tkavcik2015thermodynamics, Krotov2016DenseAM}. In biological circuits, non-equilibrium physics explains the emergence of neuronal avalanches--- i.e., bursts of activity with scale free sizes and durations--- through proximity to a critical phase transition \cite{beggs2003neuronal,Hengen2025,Cortex-Criticality2022}. The main hallmarks of near critical systems include powerlaw distributions of avalanche i) sizes $P(S)\sim S^{-\tau_s}$, ii) durations $P(D)\sim D^{-\tau_d}$, iii) average sizes versus durations $\langle S\rangle_D \sim D^{\gamma -1}$, iv) a scaling relation $\gamma \approx \frac{\tau_d - 1}{\tau_s - 1}$ connecting the three exponents, and v) universal rescalable avalanche shapes. Two decades of experiments have confirmed that cortical networks operate in quasi-critical regimes~\cite{williams2014quasicritical,fosque2021evidence} to enhance information transmission and processing\cite{shew2011information, kinouchi2006optimal,Safavi2024}. Related phenomena appear in DNNs where mean-field analyses identified an edge-of-chaos (EOC) at which signal propagation depth diverges and correlations neither collapse nor saturate \cite{schoenholz2016deep, vock2025critical, Cowsik2025}. Networks initialized near EOC display improved trainability. More recently, event-resolved analyses revealed that critical DNNs too generate scale free avalanches and exhibit a hallmarks of genuine non-equilibrium criticality~\cite{ghavasieh2025DLCriticality}. Together, these findings suggest that both biological and artificial networks may exploit similar macroscopic principles to optimize computation.

Despite the advances, our understanding of DNNs remains largely phenomenological. A rigorous mechanistic account of how such critical dynamics arises from the underlying layer-to-layer transformations is still missing. The full simulations of DNNs are expensive, and existing work has explored essentially only a single neural activation function ($\tanh$) out of many. It is therefore unknown how much of the observed critical behavior is contingent upon the specifics of architecture or activation function. An analytical framework to identify the relevant degrees of freedom, isolate the avalanche generating mechanisms, and determine possible universality classes is required.

Here I develop such a unified stochastic theory. By incorporating CLT-level fluctuations into DIP, I reduce the dynamics to a damping ratio $r$, an excitation term $h$, and two diffusion constants $D_{1}$ and $D_{2}$ controlling demographic and multiplicative stochasticity. These couplings yield a Landau description of the static critical exponents and reveal a DP–like stochastic structure to obtain the dynamic critical exponents.

I then map the neural-gain dynamics (q-dynamics) onto distinct avalanche-generating processes (x-dynamics) arising in different regions of the parameter space. The resulting x-dynamics take the form of either (i) an overdamped BM in a logarithmic trap or (ii) an absorbed free BM, corresponding respectively to a DP-like universality class with exponents $(\tau_s,\tau_d, \gamma) \approx (3/2, 2, 2)$ and a random walk (RW)-like excursion class with $(\tau_s,\tau_d, \gamma) \approx (4/3, 3/2, 3/2)$. This mapping shows how tuning the activation function steers random DNNs between universality classes.

Finally, I validate the theory numerically. Using analytically tractable toy activations, I show that the avalanche statistics agree with the universality class predictions of x-dynamics, and confirm the crossover between DP and RW scalings.

Consider a fully connected feedforward network of $N$ neurons per layer and $L$ layers, with Gaussian distributed weights $W^\ell_{ij}\sim\mathcal N(0,\sigma_w^2/N)$ and biases $b_i^\ell\sim\mathcal N(0,\sigma_b^2)$. Note that both are centered around $0$ and they have variances of $\sigma_w^2/N$ and $\sigma_b^2$ respectively.
The neural gains (pre-activations) $z^\ell$ and activations $y^{\ell+1}$ evolve as
\begin{eqnarray}
z_i^\ell = \sum_j W^\ell_{ij}y_j^\ell + b_i^\ell, \qquad y_i^{\ell+1} = \phi(z_i^\ell),
\end{eqnarray}
with $\phi(.)$ being a nonlinearity.

At the limit of $N\to\infty$, the neural gains $z_i^\ell$ become Gaussian variables with zero mean and variance $q_\ell$, yielding the deterministic recursion~\cite{schoenholz2016deep}
\begin{eqnarray}\label{eq:MF}
q_{\ell+1} = F(q_\ell) = \sigma_w^2 \int Dz \; \phi^2(\sqrt{q_\ell}z) + \sigma_b^2,
\end{eqnarray}
where $Dz = \frac{e^{-z^2/2}}{\sqrt{2\pi}} dz$ is the Gaussian measure. Equivalently, one can write $ q_{\ell+1} - q_\ell = \Delta q_\ell= \Delta F(q_\ell)$, which in the mathematical limit of continuous layers reads $\frac{dF(q_\ell)}{d\ell}$.

The deterministic map (Eq.~\ref{eq:MF}) is valid in the thermodynamic limit $\lim\limits_{N\to\infty} \frac{1}{N}\sum\limits_{i=1}^{N} \phi^2(\sqrt{q_\ell} z_i) = \int Dz \; \phi^2(\sqrt{q_\ell}z)$, where the summation samples infinite gains $z$. The actual summation $\frac{1}{N}\sum\limits_{i=1}^{N} \phi^2(\sqrt{q_\ell} z_i)$ fluctuates around its mean $\int Dz \; \phi^2(\sqrt{q_\ell}z)$. This corrects Eq.~\ref{eq:MF} for finite sizes $N$ as
\begin{eqnarray}\label{eq:MF_N_depndent}
q_{\ell+1} &=& F(q_\ell, N) = \sigma_w^2 \Big[\frac{1}{N}\sum\limits_{i=1}^{N} \phi^2(\sqrt{q_\ell} z_i) \Big] + \sigma_b^2
\end{eqnarray}
where gains are sampled from a Gaussian distribution ($z_i \sim \mathcal{N}(0,1)$) and the deterministic map is an special case: $\lim\limits_{N\to\infty}F(q_\ell, N) = F(q_\ell)$.

Note that the mean and variance of $\phi^2(\sqrt{q_\ell}z_i)$, over the probability distribution of $z_i$, read $\mu(q_\ell) = \int Dz\; \phi^2(\sqrt{q_\ell}z)$ and $v(q_\ell) = \int Dz\; \phi^4(\sqrt{q_\ell}z) - \mu(q_\ell)^2$ and the fluctuations of the finitely sampled summation follow $\mathrm{Var}[ \frac{1}{N}\sum\limits_{i=1}^{N} \phi^2(\sqrt{q_\ell}z_i) ] = \frac{1}{N^2} \sum\limits_{i=1}^{N} \mathrm{Var}[\phi^2(\sqrt{q_\ell}z_i)] = \frac{v(q_\ell)}{N}.$ This directly enables the stochastic correction of the deterministic map using CLT
\begin{eqnarray}\label{eq:Stochastic_MF}
     F(q_\ell,N)&=& F(q_\ell) + \Big(F(q_\ell,N) - F(q_\ell) \Big) \nonumber \\
     &=& F(q_\ell) + \sqrt{\frac{\sigma_w^4 \;v(q_\ell)}{N}}\eta_\ell ,
\end{eqnarray}
where $\eta_\ell \sim \mathcal{N}(0,1)$ is a Gaussian white noise. 


I Taylor expand $\phi^2(x)
= \phi_{0}^2 + \phi_{1}^{2} x + \phi_{2}^2 x^2+ \phi_{3}^2 x^3 + \phi_{4}^2 x^4 + \mathcal{O}(x^5)$ around $x=0$ and us Gaussian moments to find $\int Dz\, \phi^2\!\left(\sqrt{q}\,z\right)
= \phi_0^2 + \phi_2^2 q + 3\;\phi_4^2 \; q^2+\mathcal{O}(q^{3})$ that can be plugged into the deterministic DIP (Eq.~\ref{eq:MF}). 
    
For brevity, I define

\begin{eqnarray}
    a = -3\sigma_w^2\phi_4^2, \quad r = \sigma_w^2\phi_2^2 - 1, \quad h =\sigma_w^2\phi_0^2 + \sigma_b^2,
\end{eqnarray}

and write the solution of deterministic DIP as $q^* = \frac{r \; \pm \; \sqrt{r^2 + 4ah}}{2a}$, with $a \neq 0$ and $r^2+4ah>0$. 

Along the line of $h=0$, the absorbing state $q^* = 0$ is always a solution and the active state $q^* = r / a$ is valid only if $r/a > 0$. Along $r = 0$ the steady state follows $q^* = \sqrt{\frac{h}{a}}$ if $\frac{h}{a}\geq0$. Near critical point, the steady state behaves as
\begin{eqnarray}
    q^* &\sim r^{\beta}\qquad r \to 0,\quad h=0  \\
    q^* &\sim h^{\frac{\beta}{\sigma}} \qquad r=0, \quad h\to 0 
\end{eqnarray}

with $\beta = 1$ and $\sigma = 2$ that are known Landau static exponents, generalizing the previous derivations for $\phi = \tanh$~\cite{ghavasieh2025DLCriticality}.

Moreover, the $\phi^2$ expansion simplifies amplitude of CLT fluctuations $\frac{\sigma_w^4 \; v(q_\ell)}{N}=2D_{1}\,q_\ell \;+\; 2D_2 \;q_\ell^2$, with 
\begin{eqnarray}
    D_1=\frac{\sigma_w^4 }{2N}(\phi^2_1)^2, \qquad D_2 = \frac{\sigma_w^4 }{2N}(6\phi^2_1\phi^2_3 + 2(\phi^2_2)^2).
\end{eqnarray}

From here, it is straightforward to find the general stochastic DIP 
\begin{eqnarray}
    \Delta q_{\ell} = h + r q_\ell - a q^2_\ell + \sqrt{2D_{1}\,q_\ell \;+\; 2D_2 \;q_\ell^2} \; \; \eta_\ell,
\end{eqnarray}
whose continuous form reads: 
\begin{eqnarray}\label{eq:DP_GRW_continuous}
    \frac{dq}{d\ell} = h + r q - a q^2 + \sqrt{2D_{1}\,q \;+\; 2D_2 \;q ^2} \; \; \xi(\ell),
\end{eqnarray}
where $\xi(\ell)$ is white Gaussian noise. The equation obeys the structure of directed percolation dynamics as the canonical model for a broad family of non-equilibrium systems~\cite{diSanto2017}.

In the following, I analyze Eq.~\ref{eq:DP_GRW_continuous} across different regions of parameter space, simplify it near the absorbing state, and, via transformations of the form $q \to x$, derive the corresponding avalanche-generating dynamics. This procedure yields a unified framework for predicting DNNs' universality class. \\

Firstly, consider the case of $D_1 \gg D_2 q^{*}$, where q-dynamics simplifies to $\frac{dq}{d\ell} = h + r q + \sqrt{2 D_1  q 
} \; \; \xi(\ell)$, near the absorbing state. Using $x = 2\sqrt{q}$ transformation and Ito lemma ($d(x^2)=2xdx+(dx)^2$), I derive the x-dynamics
\begin{eqnarray}
    \frac{dx}{d\ell} = \frac{r}{2}x + \frac{2h - D_1}{x} + \sqrt{2D_1}\;\xi(\ell), \quad D_1 \gg D_2 q^{*}
\end{eqnarray}
which is an overdamped BM in the potential $U(x) = -\frac{1}{2}rx^2 - (2h-D_1)\log{x}$. At the critical point ($r\to 0$), the potential becomes a logarithmic trap $U(x) = - (2h-D_1)\log{x}$ and the excursions follow the DP-like statistics with exponents $(\tau_s, \tau_d, \gamma)\approx(3/2, 2, 2)$~\cite{LeBlanc2013,diSanto2017}. Interestingly, the same x-dynamics can produce RW-like excursions in the especial case of $h = D_1/2$~\cite{diSanto2017}.  \\

Secondly, consider the case of $\quad D_1 = 0$, where the q-dynamics follows an overdamped geometric BM $\frac{dq}{d\ell} = h + r q  + \sqrt{2D_2} \; q \; \xi(\ell), \qquad D_1 = 0,$ near the absorbing state.

The transformation $x = \log{q}$ and Ito calculus leads to the x-dynamics $\frac{dx}{d\ell} = he^{-x} -ae^{x} + (r - D_2) + \sqrt{2D_2} \; \xi(\ell)$. Taking $h=0$ it simplifies to a drifting BM 
\begin{eqnarray}\label{eq:drifting_x}
    \frac{dx}{d\ell} =&&  r - D_2 + \sqrt{2D_2}\;\xi(\ell), \\
    &&D_1 = 0,\quad  x = \log{q}.\nonumber
\end{eqnarray}
The drift vanishes $r-D_2 = 0$, with the solution of $\sigma_w^2\phi_w^2 = 1^+$, leading to a free BM, predicting an RW-like universality class, $(\tau_s, \tau_d, \gamma)\approx(4/3,3/2,3/2)$~\cite{diSanto2017}.

Note that after the transformation, the absorbing boundary moves to $x=-\infty$. So, at criticality, $x$ diffuses freely around any starting point or threshold. Let $q_{\text{th}} \ll 1$ (or equivalently $x_{\text{th}} \ll 0$) be a threshold that determines the beginning and ending of excursions. Above this threshold, deviations $x'=x-x_{\text{th}}>0$ remain small ($\sim \sqrt{2D_2}$). Since $q = e^x = q_{\text{th}} e^{x'} \approx q_{\text{th}}(1 + x')$ for small $x'$. Therefore, avalanche sizes and durations, whether measured through q- or x-dynamics, are expected to yield the same RW-like statistics.\\

Another way to confirm that free BM emerges when $D_1 =0$ is to approach the critical point from the active steady state side $q^* >0$, perturbed like $q = q^* + \varepsilon $ with $\varepsilon > 0$. The evolution governed by Eq.~\ref{eq:MF_N_depndent} follows
\begin{eqnarray}
    \frac{d\varepsilon}{d\ell} =&& (r - 2aq^*) \varepsilon +   (q^* + \varepsilon)  \sqrt{2D_2}\;\xi(\ell). \\
     &&D_1 = 0, \quad \varepsilon = q - q^*. 
\end{eqnarray}
Transforming the variable as $x = \frac{\varepsilon}{q^*}$, imposing $\varepsilon \ll q^*$, and plugging the active steady state in, reveals an overdamped BM:
\begin{eqnarray}\label{eq:over_damped_x}
    \frac{dx}{d\ell} =&& -\sqrt{r^2 + 4ah} \; x + \sqrt{2D_2} \; \xi(\ell),\\
    &&D_1 = 0,\quad x = \varepsilon/q^*. \nonumber
\end{eqnarray}

Near the critical point with $h=0$ and $r\to 0$, Eq.~\ref{eq:over_damped_x} recovers the free Brownian excursions, confirming the characterization of RW-like universality class (Eq.~\ref{eq:drifting_x}).\\

A vital question is whether the critical fluctuations survive in the thermodynamics limit ($N\to \infty$). The fluctuations magnitude can be obtained from x-dynamics (Eq.\ref{eq:over_damped_x}) near criticality. Using Ito lemma and taking the expectation value, I yield a closed ODE: $
\frac{d}{d\ell} \langle x_\ell^2\rangle = -2r\, \langle x_\ell^2\rangle + 2D_2$ with the solution
$\langle x_\ell^2\rangle = \left(\langle x_0^2\rangle - \frac{D_2}{r}\right)e^{-2r \ell} + \frac{D_2}{r}$ and the stationary variance
\begin{eqnarray}\label{eq:diffusive_susceptibility}
\lim\limits_{\ell\to\infty} \langle x_\ell^2\rangle =\langle x^2\rangle = \frac{D_2}{r}, \qquad D_1 = 0,
\end{eqnarray}
confirming that fluctuations diverge at the critical point.  

While I did not impose the absorbent boundary and reset in the above derivation, the result (Eq.~\ref{eq:diffusive_susceptibility}) is generalizable to that case. This is simply because the unconstrained x-dynamics is symmetric under $x\to -x$ and every negative valued trajectory that is counted in the calculation of $\langle x^2 \rangle$ corresponds to an equally likely positive one.

Given that the second diffusion constant scales like $D_2 \sim N^{-1}$, it is tempting to conclude that fluctuations and, consequently, avalanches vanish at the thermodynamic limit ($N\to\infty$). However, the behavior of $\langle x^2\rangle$ is not commutative at the critical point: $\Big(\lim\limits_{N\to\infty}\lim\limits_{r \to 0}\langle x^2 \rangle \neq \lim\limits_{r \to 0}\lim\limits_{N\to\infty}\langle x^2 \rangle\Big)$. To make it more concrete, let the deviation from the critical point scale like $r \sim N^{-\alpha}$ and, therefore, $\langle x^2 \rangle \sim N^{\alpha - 1}$. If $\alpha > 1$, fluctuations at the thermodynamic limit ($N\to\infty$) are amplified by critical slowing down faster than they are suppressed by self-averaging, leading to $\lim\limits_{N\to\infty}\langle x^2 \rangle = \infty$. 

Finally, I test the predictions of the theory numerically. I construct two activation functions $\phi=\Phi_{D_1}$ and $\phi=\Phi_{D_2}$ whose Taylor coefficients (Tab.~\ref{tab:tab_1}) are designed such that the theory places them in the DP-like and RW-like universality classes, respectively. I directly simulate the stochastic dynamics $q_{\ell+1}=\frac{\sigma_w^2}{N}\sum_{i=1}^{N}\phi^2(\sqrt{q_\ell}z_i)+\sigma_b^2$ (Eq.~\ref{eq:MF_N_depndent}) with $N = 1000$. They have $(q^*\approx10^{-10}, \sigma_w^2 = 1.0003)$ for $\Phi_{D_1}$ and $(q^*\approx4\times10^{-4}, \sigma_w^2 = 1.0009)$ for $\Phi_{D_2}$, and I define the threshold as $q_{\text{th}}=q^*+10^{-50}$. Each avalanche begins at $q=q_{\text{th}}$, evolves until returning below threshold, and is then reset. The duration $D$ is the number of layers above threshold, and the size is $S=\sum\limits_{\ell=1}^{D}q_\ell$. I let the full dynamics, including the reset mechanism, run for $2\times 10^{8}$ steps.

\begin{figure*}[t]
\centering
\includegraphics[width=\linewidth]{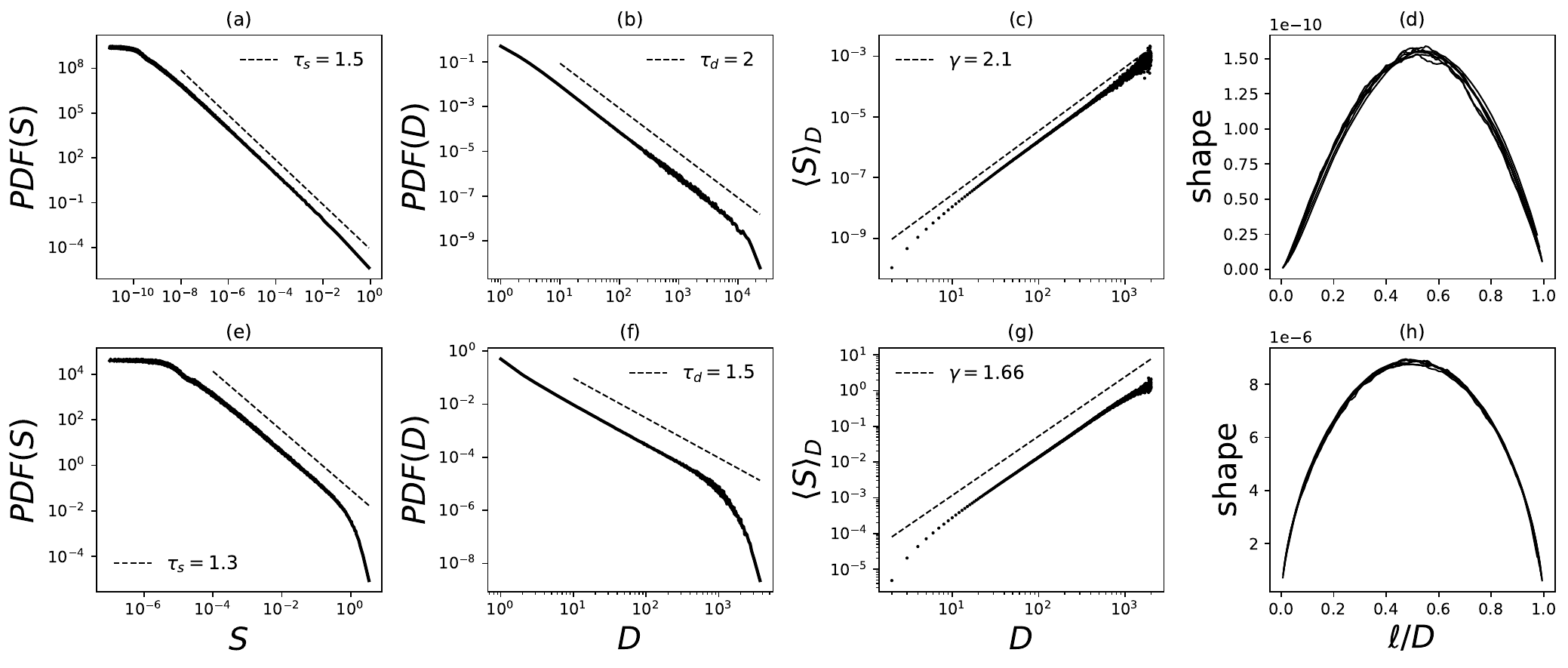}
\caption{Avalanche statistics obtained from stochastic DIP for the two activations $\Phi_{D_1}$ and $\Phi_{D_2}$ defined in Tab.~\ref{tab:tab_1}. Top row (a–d): $\Phi_{D_1}$, predicted to lie in the DP regime, shows powerlaw size $S$ and duration $D$ distributions with exponents $\tau_s \approx 3/2$ and $\tau_d \approx 2$, a crackling relation $\gamma \approx 2$ --- calculated via fitting average size $\langle S\rangle_D$ vs duration---, and a universal shape collapse. Bottom row (e–h): $\Phi_{D_2}$, predicted to yield Brownian-excursion avalanches, exhibits exponents $(\tau_s, \tau_d, \gamma) \approx (4/3, 3/2, 3/2)$ and the expected parabolic shape collapse. The results confirm that modifying the activation’s Taylor coefficients steers the universality class of random deep networks.}
\label{fig:DP_RW}
\end{figure*}

The resulting statistics (Fig.~\ref{fig:DP_RW}) display powerlaw size and duration distributions, satisfy the crackling-noise relation $\frac{\tau_d-1}{\tau_s-1}\approx\gamma$, exhibit universal shape collapse, and, crucially, match the theoretical predictions: $\Phi_{D_1}$ falls in the DP universality class with $(\tau_s,\tau_d,\gamma)\approx(3/2,2,2)$, and $\Phi_{D_2}$ realizes the RW-excursion class $(\tau_s,\tau_d,\gamma)\approx(4/3,3/2,3/2)$.

Lastly, $\Phi_{D_2}^2$ has the same Taylor coefficients as $\tanh^2$. Consequently, the present theory, supported by numerical validation, explains why previous many-body simulations~\cite{ghavasieh2025DLCriticality} reported $(\tau_s,\tau_d,\gamma)\approx(1.33,1.53,1.58)$ for $\tanh$ activation function.

\renewcommand{\arraystretch}{3}
\begin{table*}[h!]
\centering
{
\begin{tabular}{ |c | c | c | c | c | c | c | c | c | c| }
\hline
\textbf{Symbol }
& $\phi_0^2$ & $\phi_1^2$ & $\phi_2^2$ & $\phi_3^2$ & $\phi_4^2$ & $r$ & $h$ & $D_1$ & $D_2$ \\
\hline
\textbf{Definition} & $\displaystyle \phi^2(0)$ & 
$\displaystyle \frac{d\phi^2}{dx}|_{x=0}$ & 
$\displaystyle \frac{1}{2!}\frac{d^2\phi^2}{dx^2}|_{x=0}$ & 
$\displaystyle \frac{1}{3!}\frac{d^3\phi^2}{dx^3}|_{x=0}$ &
$\displaystyle \frac{1}{4!}\frac{d^4\phi^2}{dx^4}|_{x=0}$ &
$\displaystyle \sigma_w^2 \phi_2^2 - 1$ & 
$\displaystyle  \sigma_w^2 \phi_0^2 + \sigma_b^2$ & 
$\displaystyle \frac{\sigma_w^4 (\phi_1^2)^{2}}{2N}$ & 
$\displaystyle \frac{\sigma_w^4( 6\phi_1^2 \phi_3^2 + 2 (\phi_2^2)^{2})}{2N}$ \\
\hline
$\mathbf{\Phi_{D_1}}$ 
& 
$\displaystyle 10^{-20}$ & 
$\displaystyle 10^{-3}$ & 
$\displaystyle 1$ & 
$\displaystyle 0$ & 
$\displaystyle -1/2$ & 
$\displaystyle \sigma_w^2 -1$ & 
$\displaystyle 10^{-20} + \sigma_b^2$ & 
$\displaystyle\frac{10^{-6}}{2N}\sigma_w^4$ & 
$\displaystyle \frac{1}{N}\sigma_w^4$ \\
\hline
$\mathbf{\Phi_{D_2}} $
& $\displaystyle 0$
& $\displaystyle 0$
& $\displaystyle 1$
& $\displaystyle 0$
& $\displaystyle -2/3$
& $\displaystyle \sigma_w^2 - 1$
& $\displaystyle \sigma_b^2$
& $\displaystyle 0$
& $\displaystyle \frac{1}{N}\sigma_w^4$ \\
\hline
\end{tabular}
}
\caption{Taylor coefficients $\phi_n^2$ of the squared activation function $\phi^2(x)$ evaluated at $x=0$, together with the four effective couplings $(r,h,D_1,D_2)$ governing the CLT-corrected mean-field dynamics. $\Phi_{D_1}$ and $\Phi_{D_2}$ are two constructed examples that I explore in the text.}\label{tab:tab_1}
\end{table*}

In this Letter, I showed that the collective dynamics of random DNNs collapse onto a four-parameter stochastic theory with the structure of DP. This reduction explains the emergence of critical avalanches and reveals that activation functions determine the universality class. The numerical tests confirm these predictions and account for previously reported exponents~\cite{ghavasieh2025DLCriticality}. The framework provides a compact route to classifying nonequilibrium phase transitions in high-dimensional learning systems and offers a basis for engineering architectures with targeted dynamical exponents.

\bibliographystyle{apsrev4-2}
\bibliography{references} 

\appendix

\end{document}